# Asymptotic Methods of ODEs:
## Exploring Singularities of the Second Kind

**Christopher J. Winfield**

**Madison Area Science and Technology**

We develop symbolic methods of asymptotic approximations for solutions of linear ordinary differential equations and use them to stabilize numerical calculations. Our method follows classical analysis for first-order systems and higher-order scalar equations where growth behavior is expressed in terms of elementary functions. We then recast our equations in mollified form - thereby obtaining stability.

## ■ Introduction and Review

Following [1], we will study methods to develop asymptotic estimates for a system of ordinary differential equations given in the form

$$\frac{d}{dt}\overrightarrow{w(t)} = t^r A(t) \overrightarrow{w(t)} \qquad (1)$$

where $A$ is an $n \times n$ matrix with elements depending on $t$ and $\overrightarrow{w(t)}$ is a column matrix whose elements are unknown functions. We will suppose that $A$ has analytic coefficients (near infinity) and can be written as a series $A = \sum_{j=0}^{\infty} t^{-j} A_j$ (convergent for large $t$) for constant matrices $A_j$ where $A_0$ is non-trivial. (Such a singularity at finite $t$ may be turned into a singularity at infinity by a change of variables.) This type of singularity for $r > -1$ is called a *singularity of the second kind* (also known [4] as an 'irregular' singularity).

Differential equations of the form $\sum_{j=o}^{n}(t^r)^j a_j(t) \frac{d^{n-j}}{dt^{n-j}} y = 0$ ($a_j$'s bounded for large t>0 and $a_0 \equiv 1$) can also be written in the above general matrix form by setting

$$A(t) = \mathcal{A}_1 + \mathcal{A}_2 + \mathcal{A}_3 \tag{2}$$

where the $\mathcal{A}_j$'s have zero elements except for the following: $\mathcal{A}_1$ has diagonal elements 0, $-r \cdot 1\, t^{-1-r}, \ldots, -r(n-1)\, t^{-1-r}$; $\mathcal{A}_2$ has ones on the diagonal above the main diagonal; the last row of $\mathcal{A}_3$ consists of the block matrix $(-a_n(t), -a_{n-1}(t), \ldots, -a_1(t))$. Here, solutions $y$ and their first n-1 many derivatives are given by the respective components of $\vec{w}$ as $y^{(j-1)} = t^{(j-1)r} \cdot w^{(n-j-1)} : j = 1, 2, \ldots, n$.

## ■ Asymptotic Procedures

Consider a system of the form $\vec{w}' = t^r A \vec{w}$ where the elements of $A$ are analytic (near infinity) and $A \sim A_0 + t^{-1} A_1 + \ldots + t^{-r} A_{r+1}$ (formal series) where $A_0$ is a diagonal matrix for distinct complex numbers $\lambda_j$ along as the diagonal elements. Then, there are constant matrices $R$, $Q_j: j = 0, 1, \ldots, r+1$ and $P_j: j = 0, 1, \ldots$ so that the columns of

$$\Phi(t) = \left(\sum_{j=0}^{\infty} t^{-j} P_j\right) t^R \operatorname{EXP}\left(\sum_{j=0}^{r+1} t^{-j} Q_j\right) \tag{3}$$

are each asymptotic series for some solution of (1). Here the matrices $R$, $Q_j$ are each diagonal, $P_0$ may be taken as the identity matrix, and $Q_0$ may be taken to equal $A_0$; here, EXP denotes the matrix exponential and $t^R = \operatorname{EXP}(\ln t \, R)$. This is a special case of Theorems 2.1 and 4.1 of Chapt. 5 [1]. (In fact, this method produces exact solutions in cases not treated here. Cf. Chapt. 4 [1].) We will not repeat the entire proof of this result here, but we will elaborat on the construction of the $Q_j$'s and $R$. By methods which amount to a matrix version of the dominant balance method, the matrices $P_j$, $Q_{j-1}: 1 \leq j \leq r+1$ and R satisfy

$$P_k Q_0 - Q_0 P_k = \sum_{l=1}^{k} (A_l P_{k-l} - P_{k-l} Q_l) \text{ (for } 1 \leq k \leq r); \text{ and,}$$

$$P_{r+1} Q_0 - Q_0 P_{r+1} = A_{r+1} - R + \sum_{l=1}^{r} (A_l P_{r+1-l} - P_{r+1-l} Q_l)$$

For $j \geq 1$, let us denote by $\tilde{P}_j$ are the same matrix as $P_j$'s above but with zeros on their main diagonals. For our objectives, we need only to solve for the corresponding $\tilde{P}_j$'s to proceed: We obtain

$$Q_1 = \text{Diag}(A_1); \text{ and } \tilde{P}_1 Q_0 - Q_0 \tilde{P}_1 = A_1 - Q_1 \text{ off the diagonal}. \tag{4}$$

By 'Diag( )' we mean that diagonal matrix whose elements match those of the argument along the main diagonal: This is not exactly the same as the Mathematica algorithm 'Diagonal[ ]'. Then for each $2 \leq k \leq r$ (provided $r>1$) we may recursively compute the $Q_k$'s, $\tilde{P}_k$'s, and $R$ by

$$Q_k = \text{Diag}\left(A_k + \sum_{l=1}^{k-1} \left(A_l \tilde{P}_{k-l} - \tilde{P}_{k-l} Q_l\right)\right); \tag{5}$$

$$\tilde{P}_k Q_0 - Q_0 \tilde{P}_k = \sum_{l=1}^{k} \left(A_l \tilde{P}_{k-l} - \tilde{P}_{k-l} Q_l\right) \text{ off the diagonal for } 2 \leq k \leq r; \text{ and,} \tag{6}$$

$$R = \text{Diag}\left(A_{r+1} + \sum_{l=1}^{r} \left(A_l \tilde{P}_{r+1-l} - \tilde{P}_{r+1-l} Q_l\right)\right). \tag{7}$$

## ■ Example System of Equations

We begin with an example involving rational functions of t. (Such examples suffice in our general setting by our hypothesis on $A(t)$). Consider the case $r=1$, $x(0) = 0$, $y(0) = 4$, and

$$A(t) = \begin{pmatrix} -0.1 & 0 \\ 0 & -0.5 \end{pmatrix} + t^{-1}\begin{pmatrix} 1 & 1 \\ 1 & 0 \end{pmatrix} + t^{-2}\begin{pmatrix} 0 & -1 \\ 0 & -1 \end{pmatrix}. \text{ We set} \tag{8}$$

```
In[1]:= A0 = {{-1, 0}, {0, -0.5}}; Q0 = A0; P0 = IdentityMatrix[2];
        A1 = {{1, 1}, {1, 0}};
```

as we need only terms of order $t$ and 1 for accuracy up to $O(t^{-1})$ in the first factor of (3). We follow equations (4) to solve $Q_1$ and $\tilde{P}_1$ (we will denote respective $\tilde{P}_j$'s by Pj's in our inputs cells) and (7) to solve $R$:

```
In[2]:= preP1 = Array[a, {2, 2}]; NN = 2;
        C1 = A1 - DiagonalMatrix[Diagonal[A1]];
        E1 = preP1 . Q0 - Q0 . preP1;
```

```
av1[j_, m_] :=
 If[j == m, 0,
  a[j, m] /.
    Flatten[
     Solve[Flatten[Table[E1[[k, l]] == C1[[k, l]],
        {k, NN}, {l, NN}]],
   Delete[Flatten[Table[a[k, l], {k, NN}, {l, NN}]],
       Array[{1 + (NN + 1) * (#1 - 1)} & , NN]]]]]
P1 = Simplify[Array[av1, {NN, NN}]];
Q1 = Simplify[DiagonalMatrix[Diagonal[A1]]];
PreR = A1.P1 - P1.Q1;
R = DiagonalMatrix[Diagonal[PreR]];
```

Asymptotic solutions up to first derivatives are given by the columns of the matrix *V* which we compute by

```
In[7]:= V = Simplify[(P0 + P1 / t).MatrixExp[Log[t] R].
     MatrixExp[ t^2 Q0 / 2 + t * Q1]];
    MatrixForm[V]
```

*Out[7]//MatrixForm=*

$$\begin{pmatrix} 0. + \frac{e^{t-\frac{t^2}{2}}}{t^{2.}} & 2. \, e^{-0.25\, t^2}\, t^{1.} \\ -\frac{2.\, e^{t-\frac{t^2}{2}}}{t^{3.}} & 0. + e^{-0.25\, t^2}\, t^{2.} \end{pmatrix}$$

We find that there are two linearly independent solution vectors $\vec{V}_1$ and $\vec{V}_2$ satisfying

$$\vec{V}_1 = e^{-0.25\, t^2}\begin{pmatrix} 2\, t(1 + o(t)) \\ t^2(1 + o(t)) \end{pmatrix};\ \vec{V}_1 = e^{t-\frac{t^2}{2}}\begin{pmatrix} t^{-2}(1 + o(t)) \\ -2\, t^{-3}(1 + o(t)) \end{pmatrix} \text{ as } t \longrightarrow +\infty.$$

We compare our result with exact solution of an initial value problem of the following asymptotically equivalent [2] system:

$$\frac{d}{dt}\overrightarrow{x(t)} = \begin{pmatrix} 1-t & 1 \\ 1 & -t/2 \end{pmatrix}\overrightarrow{x(t)};\ \overrightarrow{x(0)} = \begin{pmatrix} 1 \\ 1 \end{pmatrix}$$

Routine calculations, involving hypergeometric functions, show that $\overrightarrow{x(t)} = C\, e^{-0.25\, t^2}\begin{pmatrix} 2\, t(1 + o(t)) \\ t^2(1 + o(t)) \end{pmatrix}$ for a constant $C \approx 12.09$.

## ◻ Instability

A natural way to check our results is to compute numerical results via NDSolve[ ]. Here, we compare our dominant asymptotic calculations (from $\vec{V}_2$) to numerical results (via Interpolation) of the original equation. Below we expect our graphs to have horizontal asymptotes; instead, we find the calculations to be unstable for $t > 9$.

```
In[8]:= F[t] := -1/t^2 + 1/t;
    S = NDSolve[{x'[t] == (1 - t) * x[t] + t * F[t] * y[t],
        y'[t] == -.5 t y[t] + t * F[t] * x[t], x[2] == 5, y[2] == 5},
        {x, y}, {t, 2, 11}];
    Testx = x[t] / V[[1, 2]] //. S;
    Testy = y[t] / V[[2, 2]] //. S;
    Plot[{Testx, Testy}, {t, 2, 11}, PlotRange → Automatic,
        AxesLabel → {t}]
```

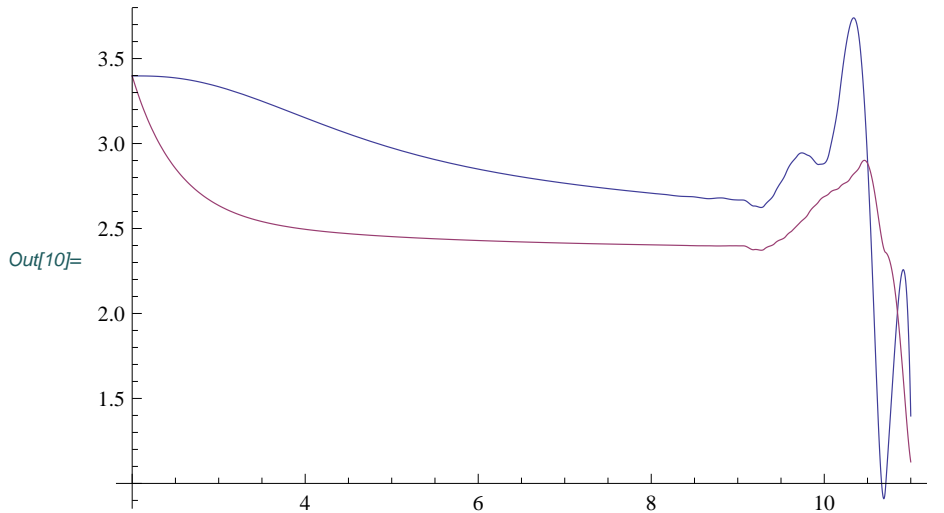

## ◻ Mollification

We can apply these asymptotic estimates to produce stable solutions from NDSolve[ ] for large $t$. We will replace $x(t)$ by $2\,t e^{-.25\,t^2}\,x(t)$ and $y(t)$ by $t^2 e^{-.25\,t^2}\,y(t)$ to produce a solution $(x, y)$ with less drastic decay rates:

```
In[11]:= newsys =
          {D[x[t] (t + 1) Exp[-.25 t^2], t] ==
              (1 - t) (t + 1) Exp[-.25 t^2] x[t] +
               (t + 1)^2 Exp[-.25 t^2] y[t],
            D[Exp[-.25 t^2] (1 + t)^2 y[t], t] ==
              (t + 1) Exp[-.25 t^2] x[t] -
               0.5 t Exp[-.25 t^2] (1 + t)^2 y[t]};
        Reformedprob =
          Flatten[NDSolve[{newsys, x[0] == 1, y[0] == 1},
             {x[t], y[t]}, {t, 0, 300}]];
        Plot[{x[t] //. Reformedprob, y[t] //. Reformedprob},
         {t, 0, 600}, AxesLabel → {t}]
```

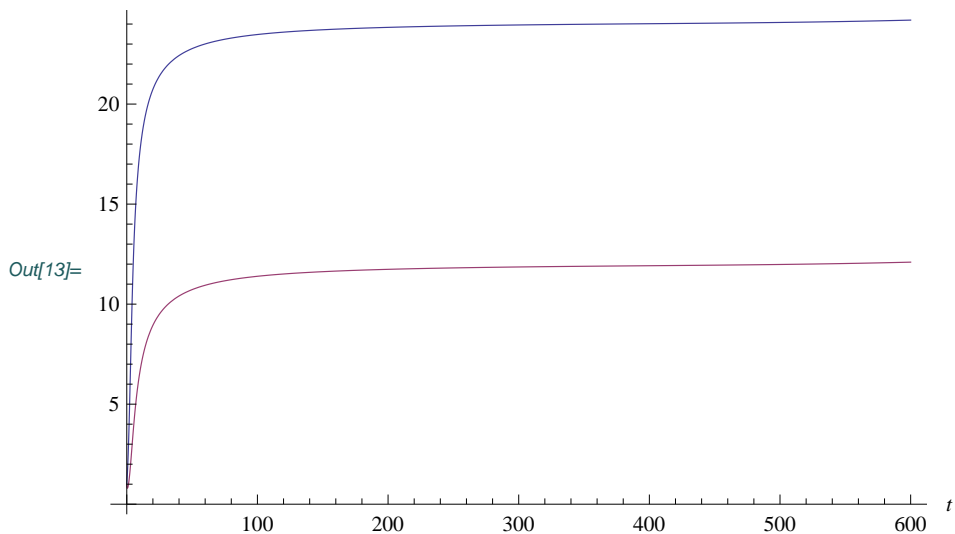

The solutions seem to be tending to limiting values for *t* near 600 : All with no call to any special calculational options.

It is beyond the scope of this article to compare and contrast numerical schemes since our objective is to predict theoretically the growth behavior of solution as demonstrated using default options of NDSolve[ ]. Our method, we note, may serve to improve stability and reduce stiffness, yet we do not quantify such results. We do, however, note that widely varying exponential growth/decay of solutions are known to produce instability similar to those we see (References [6], [7], [8] and NDSolve of the Documentation Center [9], among many others, introduce rigorous methods regarding such issues.)

## ■ Application to a Third-order Ordinary Differential Equation

We will study asymptotic behavior of solutions to $L\, y = 0$ ($t > 0$) for the operator $L = D^3 + (t^2 - 3\,t)\,D^2 - t^4\,D$. Following (2) we will study the system $\vec{w}\,' = M\,\vec{w}$ which we can write as

$$M = t^2 \left[ \begin{pmatrix} 0 & 1 & 0 \\ 0 & 0 & 1 \\ 0 & 2 & -1 \end{pmatrix} + \frac{1}{t}\begin{pmatrix} 0 & 0 & 0 \\ 0 & 0 & 0 \\ 0 & 0 & 3 \end{pmatrix} + \frac{1}{t^2}\begin{pmatrix} 0 & 0 & 0 \\ 0 & 0 & 0 \\ 0 & 0 & 0 \end{pmatrix} + \frac{1}{t^3}\begin{pmatrix} 0 & 0 & 0 \\ 0 & -2 & 0 \\ 0 & 0 & -4 \end{pmatrix} \right] \qquad (9)$$

We choose this example so as to force a diagonalization step before applying the procedure to produce estimates (3), after which we identify $A_j$: $j = 0, 1, 2, 3$ (with $r = 2$)

```
In[14]:= NN = 3; rr = 2; M0 = {{0, 1, 0}, {0, 0, 1}, {0, 2, -1}};
        M1 = {{0, 0, 0}, {0, 0, 0}, {0, 0, 3}};
        M2 = {{0, 0, 0}, {0, 0, 0}, {0, 0, 0}};
        M3 = DiagonalMatrix[Array[-(# - 1) * rr &, NN]];
```

We proceed with the diagonalization step:

```
In[15]:= Clear[TT]; TT = Transpose[Eigenvectors[M0]];
        A0 = Inverse[TT].M0.TT; A1 = Inverse[TT].M1.TT;
        A2 = Inverse[TT].M2.TT;
        A3 = Inverse[TT].M3.TT;
```

Here $Q_0$, and $Q_1$ are obvious and, using (4) and (5), we compute $\tilde{P}_1$ and $Q_2$:

```
In[17]:= Q0 = A0; Q1 = DiagonalMatrix[Diagonal[A1]];
        preP1 = Array[a, {3, 3}];
        EE10 = preP1.Q0 - Q0.preP1; EE20 = preP2.Q0 - Q0.preP2;
        CC1 = A1 - Q1;
        av1[j_, m_] :=
          If[j == m, 0,
            a[j, m] /.
              Flatten[
                Solve[Flatten[Table[EE10[[k, l]] == CC1[[k, l]],
                    {k, NN}, {l, NN}]],
                  Delete[Flatten[Table[a[k, l], {k, NN}, {l, NN}]],
                    Array[{1 + 4 * (# - 1)} &, NN]]]]];
        P1 = Array[av1, {NN, NN}]; PreQ2 = (A1 - Q1).P1 + A2;
        Q2 = DiagonalMatrix[Diagonal[PreQ2]];
```

Using (6) and (7) we compute $\tilde{P}_2$ and $R$:

```
preP2 = Array[aa, {3, 3}];
CC21 = A2 - Q2 - DiagonalMatrix[Diagonal[A2 - Q2]];
CC22 = A1.P1 - P1.Q1 - DiagonalMatrix[Diagonal[A1.P1 - P1.Q1]];
av2[j_, m_] :=
  If[j == m, 0,
    aa[j, m] /.
      Flatten[
        Solve[
          Flatten[
            Table[EE20[[k, l]] == CC21[[k, l]] + CC22[[k, l]],
              {k, NN}, {l, NN}]],
          Delete[Flatten[Table[aa[k, l], {k, NN}, {l, NN}]],
            Array[{1 + 4 * (# - 1)} &, NN]]]]];
P2 = Array[av2, {NN, NN}]; PreR = (A1 - Q1).P2 + (A2 - Q2).P1 + A3;
R = DiagonalMatrix[Diagonal[PreR]];
```

We are ready to compute asymptotic solutions after a of change basis:

*In[25]:=*
```
preV =
  Simplify[MatrixExp[Log[t] R].
    MatrixExp[ t^3 Q0 / 3 + t^2 Q1 / 2 + t Q2]];
V = Simplify[TT.preV];
```

Now, we multiply rows of the solution matrix by various power of *t* to produce the various asymptotic estimates of solutions to $L\, y = 0$ which we can read off from the columns of "Soln".

*In[27]:=*
```
S = DiagonalMatrix[Array[t^(rr (#-1)) &, NN]]; Soln = S.V;
MatrixForm[Soln]
```

*Out[27]//MatrixForm=*

$$\begin{pmatrix} \dfrac{e^{-\frac{1}{3} t \left(2 - 3 t + 2 t^2\right)}}{t^{32/9}} & \dfrac{e^{\frac{1}{6} t \left(4 + 3 t + 2 t^2\right)}}{t^{22/9}} & 1 \\ -\dfrac{2\, e^{-\frac{1}{3} t \left(2 - 3 t + 2 t^2\right)}}{t^{14/9}} & \dfrac{e^{\frac{1}{6} t \left(4 + 3 t + 2 t^2\right)}}{t^{4/9}} & 0 \\ 4\, e^{-\frac{1}{3} t \left(2 - 3 t + 2 t^2\right)} t^{4/9} & e^{\frac{1}{6} t \left(4 + 3 t + 2 t^2\right)} t^{14/9} & 0 \end{pmatrix}$$

We find three different types of behavior for asymptotic solutions: rapid decay, rapid growth, and a constant solution. We can check that any constant function is indeed a solution, so the constant behavior is precise and not just asymptotic in this case. Moreover, any solution *y* is majorized by Soln[[1, 2]] = $t^{-22/9}\, e^{\frac{1}{6} t (4 + 3 t + 2 t^2)}$.

## ☐ **Graphical Results**

We check our results using two approaches. First we simply graph ratios $\vec{w}/\text{Soln}[[2,1]]$ with numerical solutions $\vec{w}$. Let us consider this using a non-constant solution arising from the initial values $(w(0), w'(0), w''(0))=(0,1,0)$ (the analysis for initial values $(0, 0, 1)$ is similar and we invite the reader to check):

```
In[28]:= diffop = y'''[t] - 2 t^4 y'[t] - (-t^2 + 3 t) y''[t];
        initvprob =
         NDSolve[{diffop == 0, y[0] == 0, y'[0] == 1, y''[0] == 0},
          y[t], {t, 0, 15}];
        Plot[(y[t] / Soln[[1, 2]]) /. initvprob, {t, 0, 15},
         PlotRange → All, AxesLabel → {t, ratio}]
```

*Out[29]=*

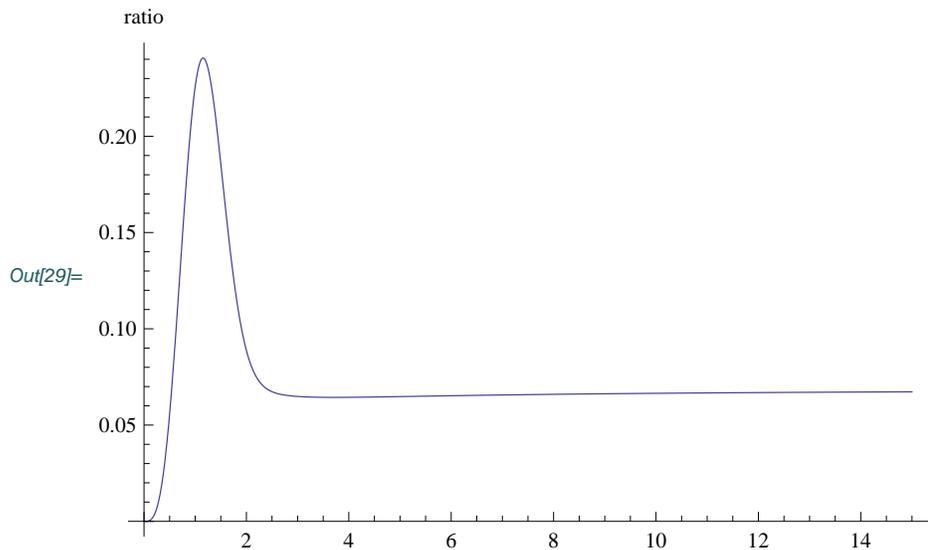

We are able to produce graphs using default options of NDSolve[ ] only for moderate domains of *t*. For our second approach we mollify the operator by replacing $y(t) \longrightarrow y(t) \times t^{-22/9} e^{\frac{1}{6} t (4 + 3t + 2t^2)}$, thereby theoretically guaranteeing bounded solutions *y* on intervals away from the origin $t = 0$.

```
In[30]:= newdiffop =
    Simplify[
      (D[y[t] * Soln[[1, 2]], {t, 3}] -
         2 t^4 D[y[t] * Soln[[1, 2]], {t, 1}] +
         (t^2 - 3 t) D[y[t] * Soln[[1, 2]], {t, 2}]) / Soln[[1, 2]]];
    t0 = 30;
    Mollified =
     NDSolve[{newdiffop == 0, y[t0] == 1, y'[t0] == 0, y''[t0] == 0},
      y[t], {t, t0, t0 + 200}];
    Plot[y[t] /. Mollified, {t, t0, t0 + 200}, AxesLabel → {t, y}]
```

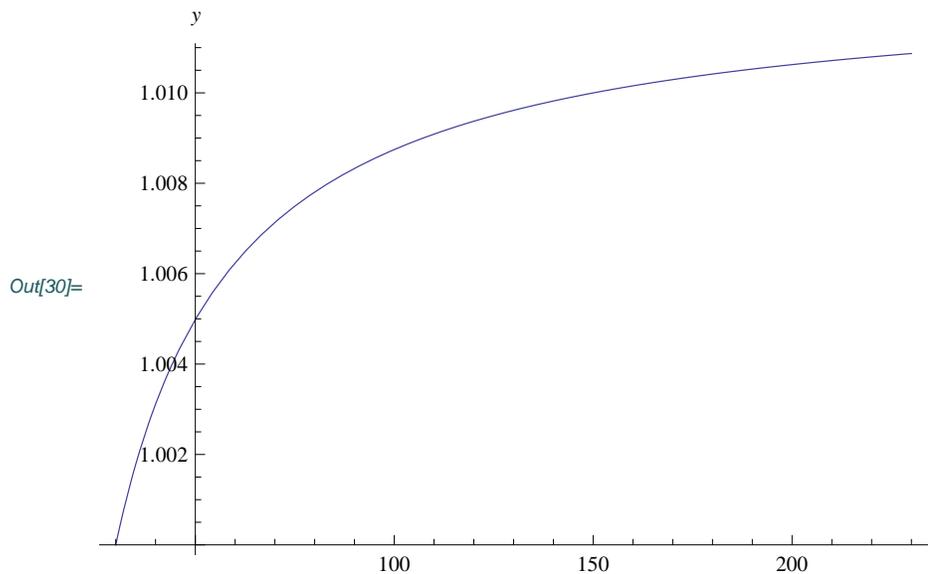

We find instability as we calculate relative errors over large intervals (see [5]) likewise to those of the previous section.

Alternatively, we use a mollified operator along with a linear change of variable to calculate solutions on intervals for large *t*. We apply the transformation $t \rightarrow \tau + q$ after applying the mollification procedure above: Then the initial value problem applied near $t = 0$ for the new problem is equivalent to solving the original near $t = q$ for large *q*. We find that in this way we can accurately compute solutions on large intervals in *t*:

```
In[31]:= Manipulate[newfactor = Soln[[1, 2]] //. t → τ + q;
         newdiffop =
          Simplify[
           (D[y[τ] * newfactor, {τ, 3}] -
              2 (τ + q)^4 D[y[τ] * newfactor, {τ, 1}] +
              ((τ + q)^2 - 3 (τ + q)) D[y[τ] * newfactor, {τ, 2}]) /
            newfactor];
         newsol =
          NDSolve[{newdiffop == 0, y[0] == 10, y'[0] == 0, y''[0] == 0},
           y[τ], {τ, 2500}];
         Plot[y[τ] /. newsol, {τ, 0, 200}, AxesLabel → {τ, y}],
         {q, 400, 500}]
```

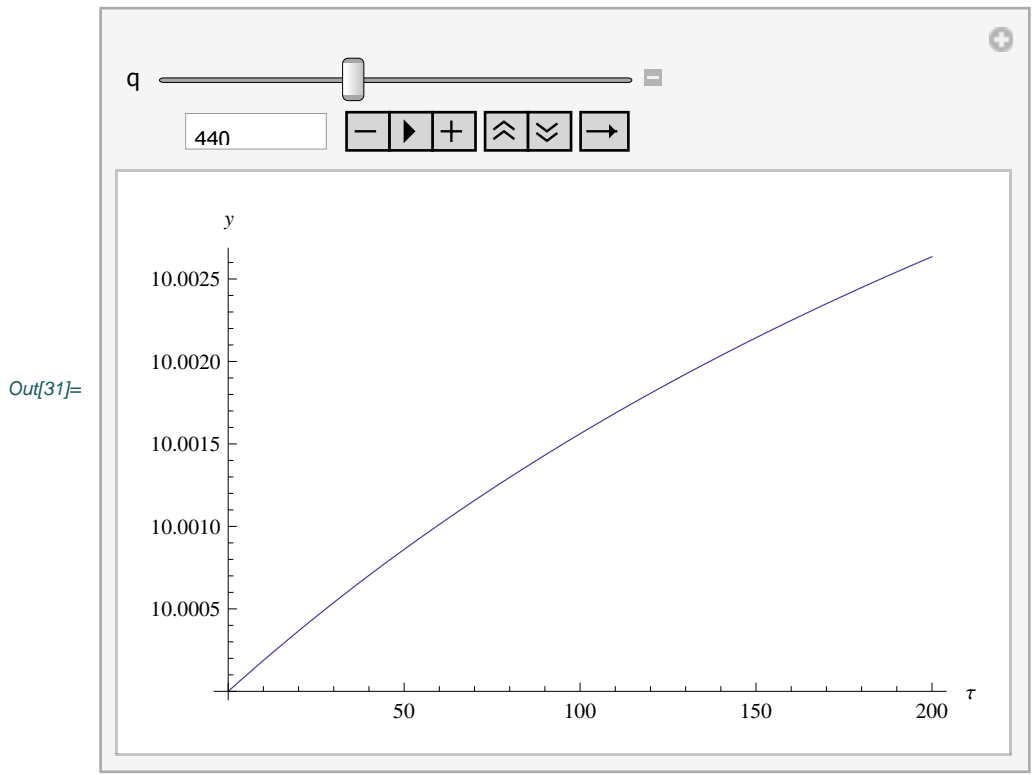

Out[31]=

We note that the advantage here is that our end result is of the form $y(t) \times t^{-22/9} e^{\frac{1}{6} t (4+3 t+2 t^2)}$ for large *t* where *y* is computed with considerable accuracy and the other terms are known exactly.

# ■ Conclusions

We find that classical techniques of asymptotics can be calculated symbolically and in an automated way via simple matrix manipulations. These estimates can serve to check accuracy of numerical solutions both for scalar equations and for systems of equations of the types studied. Moreover, these estimates may serve to transform differential equations to certain mollified versions which admit solutions bounded as $t \longrightarrow +\infty$: As solutions of mollified problems can be computed with accuracy and stability, with the transformation in elementary form, accuracy in our calculations is improved overall.

## ■ Acknowledgments

The author would like to thank the members of Madison Area Science and Technology for their discussions of Mathematica programming and various applications.

## ■ References

### About the Author

Dr. Winfield holds a PhD in Mathematics/Real Analysis from UCLA (1996) and an MS in Physics from UW-Madison (1989). He has publications in partial differential equations, quantum scattering theory and cosmology/Einstein's equation. He currently works as consultant with the Madison Area Science and Technology group based in Madison WI.

**Christopher Winfield**
*cjwinfield@madscitech.org*